\documentclass{aa}
\usepackage{graphicx}
\def\approxgt{\mathrel{\hbox{\rlap{\lower.55ex \hbox {$\sim$}}
        \kern-.3em \raise.4ex \hbox{$>$}}}}
\def\approxlt{\mathrel{\hbox{\rlap{\lower.55ex \hbox {$\sim$}}
        \kern-.3em \raise.4ex \hbox{$<$}}}}

\begin{document}
   \title{Drilled by the jet? XMM-Newton discovers  \\ a 
Compton-thick AGN in the GPS galaxy Mkn~668}

   \author{Matteo Guainazzi
          \inst{1},
	  Aneta Siemiginowska
	  \inst{2},
	  Pedro Rodriguez-Pascual
	  \inst{1},
	  Carlo Stanghellini
	  \inst{3}
          }

   \offprints{M.Guainazzi}

   \institute{$^1$XMM-Newton Science Operation Center, VILSPA, ESA, Apartado
              50727, E-28080 Madrid, Spain \\
              \email{mguainaz@xmm.vilspa.esa.es} \\
	      $^2$Harvard-Smithsonian Center for Astrophysics, 60 Garden St.,
		Cambridge, MA 02138, USA \\
 	      $^3$Istituto di Radioastronomia CNR, Noto, Italy \\
              }

   \date{Received ; accepted }

   \abstract{
We report the XMM-Newton discovery of the first
Compton-thick obscured AGN in a Broad Line Radio Galaxy, the 
Gigahertz Peaked-Spectrum source Mkn~668 (OQ+208). The
remarkably flat 2--10~keV X-ray spectrum (observed
photon index, $\Gamma \simeq$ 0.7), alongside with a prominent
iron K$_{\alpha}$ fluorescent emission line, is a clear signature of
a Compton-reflection dominated spectrum. Mkn~688 represents a remarkable example of discrepancy between
X-ray spectral properties and optical classification,
as its optical spectrum is characterized by broad and asymmetric
Balmer lines.
The obscuring matter
is constrained to be located within the radio hotspots, in turn
separated by about 10~pc. If the jets are piercing their
way through a Compton-thick medium pervading the
nuclear environment, one could be largely underestimating
the radio activity dynamical age determined
from the observed hotspot recession velocity.
The soft X-ray spectrum is dominated by a much
steeper component, which may be due to nuclear continuum electron scattering,
or inverse Compton of the -
remarkably large - far infrared emission.
Soft X-rays are suppressed by a
further Compton-thin
($N_H \sim 10^{21}$~cm$^{-2}$) absorbing system, that we
identify with
matter responsible for free-free absorption of the radio lobes.
            }

\authorrunning{Guainazzi et al.}

\titlerunning{The Compton-thick AGN in the GPS galaxy Mkn~668}

   \maketitle
%

\section{Introduction}

Gigahertz Peaked-Spectrum (GPS) radio sources
are a class of compact (10--100~mas; 10--100~parsecs)
radio sources, characterized by a simple convex radio 
spectrum peaking around 1~GHz (see O'Dea et al. 1998
for a review). Together with the (despite of their name)
less compact and steeper
Compact Steep-Spectrum (CSS) sources, they constitute
a sizable fraction of the 5~GHz selected sources,
probably as large as 40\% (\cite{odea98}).

In the ROSAT All Sky Survey,
GPS/CSS quasars
exhibited a detection rate 3 times lower than
radio-loud quasars of comparable power
(\cite{baker95}). The explanation for
their X-ray weakness is still matter of debate.
They
may represent an intrinsically X-ray weak population
of radio-loud Active Galactic Nuclei (AGN); 
alternatively, their flux in the soft
X-ray band may be suppressed by matter heavily obscuring
the active nucleus. The first pointed
observations with ROSAT suggested
that X-ray absorption in GPS/CSS quasars may be
common (\cite{elvis94}).
The first hard X-ray measurements of GPS galaxies
needed to await ASCA, which unveiled column densities of
the order of $10^{22}$~cm$^{-2}$ (\cite{odea00},
\cite{guainazzi00}). However, the number of
GPS galaxies for which X-ray spectroscopic measurements
are available remains to-date dramatically low.

The 
X-ray properties of GPS galaxies may have immediate
consequences on our understanding of their nature,
and, in turn, on the evolution
of large-scale radio structures in the universe.
The detection of hotspot proper motions
in an handful of Compact Symmetric Objects (\cite{polatidis03}),
corresponding to dynamical ages $\sim 10^2$--$10^3$~years,
suggests that at least some GPS galaxies should
represent an infancy stage of radio galaxies proper
(\cite{phillips82}, \cite{carvalho85},
\cite{fanti95}, \cite{readhead96}).
However, there is convincing
evidence for the
presence of large amount of matter in the core
of GPS sources (\cite{devries98}, \cite{pihlstrom03},
\cite{snellen02}). It is still
unclear whether this matter may have
the average density ($\sim 1$--10~cm$^{-3}$) required to 
ensure the permanent confinement of the radio
structure (\cite{deyoung93}), and to
prevent (``frustrate'') its
full development into a Fairnoff-Railey (FR)~II,
or, most likely, into a FR~I galaxy (\cite{odea00}). X-ray
probing of the circumnuclear environment
may represent an important clue.

Thanks to the unprecedented collecting area of
its optics, XMM-Newton (\cite{jansen01})
is the ideal observatory to perform high-sensitivity,
moderate-resolution spectroscopy of weak
($10^{-13}$~erg~cm$^{-2}$~s$^{-1}$) X-ray sources. An
observation program is being carried on,
aiming at studying the properties of
the circumnuclear gas in a sizable sample
of GPS galaxies. In this paper, we report
on the observation of Mkn~668, which
constitutes to date one of the very few hard X-ray
measurements of a GPS galaxy ever.
XMM-Newton
unveiled the first, to our knowledge, Compton-thick AGN
in a broad-line radio galaxy.
Mkn~668 (OQ+208, 1404+286) is one
of the closest known GPS galaxy
($z=0.0077$; \cite{stanghellini93}).
The host galaxy shows signs of tidal distortion
(\cite{stanghellini93}), suggesting
a recent merging event. Optically classified as
a Seyfert~1 (\cite{blake70}, \cite{marziani93}),
its radio morphology exhibits two radio lobes
along the NE:SW direction at the approximate distance of 10~pc.
Marginal evidence for a radio core closer to
the former and brighter lobe were recently
reported by Lister (2003).
The whole radio structure is embedded in a large
diffuse halo of diameter $\simeq$30~kpc (\cite{debruyn90}).
The
hotspots separate with an apparent velocity of
30--60$\mu$as~yr$^{-1}$ (\cite{stanghellini02},
\cite{lister03}, \cite{barone03}). This identifies
Mkn~668 as a potential ``young'' radio source, with
an estimated dynamical age of
$\simeq$100--200~years. Kameno et al. (2000)
fit the
long-wavelength radio cutoff with free-free
absorption by matter embedding the radio
structure. They claimed that
thermal emission from the absorbing gas
is responsible for the soft X-ray emission detected by 
ROSAT (Zhang \& Marscher 1994). 

An ASCA observation of Mkn~668
(\cite{guainazzi03}) unveiled a very
flat X-ray spectrum, together with a possible
bright (Equivalent Width, $EW \simeq 900$~eV)
K$_{\alpha}$ fluorescent iron line. These
results
prompted the XMM-Newton observation, which is the main
topic of this paper.

In this paper: energies are quoted in the source frame;
errors are quoted at the 1-$\sigma$ level for the
count rates, and at the 90\% confidence level for 1
interesting parameter for fit parameters
and derived quantities; $H_{\rm 0} = 70$~km~s$^{-1}$~Mpc$^{-1}$;
and $q_0 = 0.5$. At the Mkn~668 redshift,
1$\arcsec \simeq 1.5$~kpc.

\section{XMM-Newton observation}

XMM-Newton
observed the sky region around Mkn~668 on
January 31, 2003. In this paper only results from the
EPIC cameras (pn; \cite{struder01}; MOS, \cite{turner01})
will be discussed, as Mkn~668 is too weak
to be detected by the high-resolution
spectroscopy cameras (RGS). Data were reduced with
{\sc SAS v5.4.1} (\cite{jansen01}), using the most updated
calibration files available at the moment the reduction
was performed (March 2003). Standard procedures
were followed for the EPIC data reduction and analysis,
as detailed in the {\it SAS User's Guide} (\cite{loiseau03}).
In particular: scientific products were accumulated combining
single and double (to quadruple) pn (MOS)
events (this is most appropriate for
moderately weak sources to enhance the signal-to-noise ratio);
background products were extracted from regions belonging to the same
CCD as Mkn~668 and free form contaminating
sources; intervals of high particle
background were removed, applying standard thresholds to the
$E>10$~keV, single event field-of-view light curves
(1 and 0.35 counts per second for the pn and the MOS
cameras, respectively).
After screening, the total exposure time is
13.2 and 15.5~ks for the pn and MOS cameras, respectively.
The event lists of the two MOS cameras were
merged together before extracting any scientific products. 
Version 5.1 of the {\sc LHEASOFT}
package was employed
for the scientific analysis described hereinafter.

\subsection{Imaging}

In the EPIC field-of-view image
a source is clearly detected close to the nominal
boresight positions. In the pn,
$\alpha_{{\rm 2000}}=14^h07^m00.3^s$,
$\delta_{{\rm 2000}}=+28^{\circ}27\arcmin 14\arcsec$, The
0.5--10~keV count rates are $(8.6 \pm 0.4)$, and
$(6.0 \pm 0.3) \pm 10^{-2}$~s$^{-1}$ in the pn and
MOS cameras, respectively.
The best-fit centroid is 1.5$\arcsec$ distant from the galaxy
core optical position, consistent with typical
XMM-Newton
attitude reconstruction accuracy.
There is no evidence for intrinsic extension beyond the moderately
broad XMM-Newton mirror
Point Spread Function ($6\arcsec$ Full Width Half Maximum; FWHM).
Two additional sources are
detected at a signal-to-noise ratio larger than 5 within 3$\arcmin$
from the Mrk~668 nucleus ({\it i.e.} within the
typical ASCA aperture; see Sect.~3.1). Their position and count rates
are reported in Tab.~\ref{tab2}.
\begin{table}
\caption{Position and count rate
($CR$) of the sources detected by the XMM-Newton EPIC cameras
within 3$\arcmin$ from the optical nucleus
of Mrk~668, and having a
signal-to-noise ratio larger then 5}
\begin{center}
\begin{tabular}{lccc} \hline \hline
$\alpha_{{\rm 2000}}$ & $\delta_{{\rm 2000}}$ & 0.5--10~keV & \\
& & CR (10$^{-2}$s$^{-1}$) \\ \hline
$14^h07^m00.3^s$$^a$ & $+28^{\circ}27\arcmin 14\arcsec$$^a$ & $14.5 \pm 0.4$$^a$ & \\
$14^h06^m57.4^s$ & $+28^{\circ}28\arcmin 5\arcsec$ & $1.41 \pm 0.19$ & \\
$14^h07^m04.4^s$ & $+28^{\circ}29\arcmin 37\arcsec$ & $1.03 \pm 0.18$ \\
\hline \hline
\end{tabular}
\end{center}

\noindent
$^a$Mkn~668

\label{tab2}
\end{table}
The sum of their count rates
amounts at $29\pm10\%$ and $<11\%$
of Mrk~668 count rate in the 0.5--2 and 2--10~keV energy
bands, respectively.

\subsection{Spectral analysis of the nuclear emission}

\begin{figure}
   \centering
   \includegraphics[angle=-90,width=8cm]{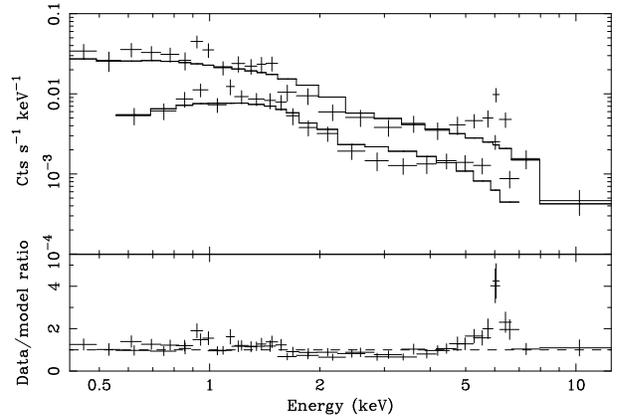}
      \caption{Spectra ({\it upper panel}) and
		residuals in units of data/model
		ratio ({\it lower panel})
		when a photoelectrically absorbed power-
		law model is simultaneously fit to the MOS and
		pn spectra of Mkn~668. Each data point corresponds
		to a signal-to-noise ratio larger than 3
              }
         \label{fig7}
   \end{figure}

\begin{figure}
   \centering
   \includegraphics[angle=-90,width=8cm]{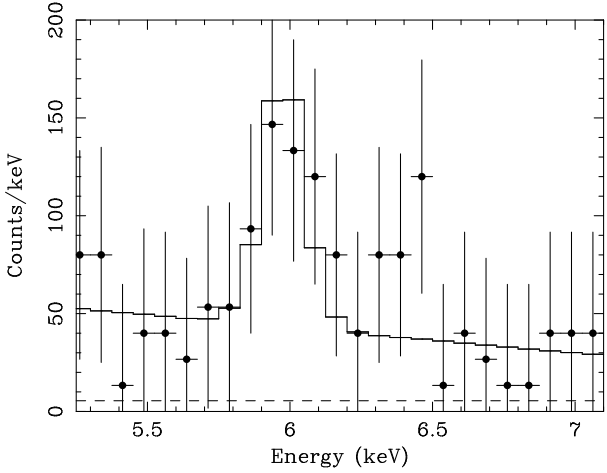}
      \caption{pn non-background subtracted spectrum in
	      the 5.25--7.25~keV energy band. Data points have
		been rebinned with a
		constant width of 75~eV (which
		corresponds to about one-third of the
		intrinsic energy resolution). The {\it
		solid line} represents a fit with
		a power-law continuum and an
		unresolved Gaussian profile. The
		background level is indicated by
		the {\it dashed line}
              }
         \label{fig4}
   \end{figure}

Mkn~668 nucleus spectra and light curves were extracted
from regions of 40$\arcsec$ and 25$\arcsec$ radius for the pn
and the MOS, respectively.
Total source counts are $880 \pm 30$ and $620 \pm 20$
in the pn and MOS, respectively, 15\% and 11\% of which
are due to background.
No clear evidence for variability of the Mkn~668 nucleus flux
is seen during the XMM-Newton observation.
A fit with a constant function on the pn 1024~s binned
light curves in the 0.5--2~keV and 2--10~keV energy bands
yields $\chi^2_{\nu} = 0.81$ and $\chi^2_{\nu} = 1.01$,
respectively. We will therefore focus in the following
on the time-averaged spectra
only.
Spectra were binned in order to have
at least 25 counts in each spectral channel, and to oversample
the intrinsic energy resolution of each EPIC camera by
a factor not larger than 3.
Spectral fits were performed in the 0.35--15~keV and
0.5--10~keV energy ranges for the pn and MOS, respectively
A fit with
a one-component model\footnote{All the models discussed
in this paper are photoabsorbed by cold matter,
whose column density is held fixed to the contribution
of our Galaxy along the line-of-sight to Mkn~668:
$N_{{\rm H,Gal}} = 1.4 \times 10^{20}$~cm$^{-2}$ (\cite{dickey90})} is
clearly inadequate (if a power-law
model is used: $\chi^2=111.0/47$~$\nu$, where
$\nu$ indicates the number of degrees of freedom;
the residuals against this fit are shown in Fig.~\ref{fig7}).
Phenomenologically, a
good fit can be obtained with
a two continuum model such a broken power-law
(cf. Tab.~\ref{tab2}), provided
that a Gaussian emission line is included to account for
a narrow-band feature around 6~keV (observer's frame).
\begin{table*}
\caption{Best-fit parameters and results for global fits to the
nuclear spectra of Mkn~668. Models legenda: \#1~=~BKNPO+GA; \#2~=~WA$^{soft}$$\times$(PO$^{soft}$+2$\times$GA)+WA$^{hard}$$\times$(PO$^{hard}$+GA); \#3~=~WA$\times$(PO$^{soft}$+PEXRAV+3$\times$GA); \#4~=~WA$\times$(MEKAL+PEXRAV+GA), where: PO~=~power-law, BKNPO~=~broken power-law; PEXRAV~=~Compton-reflection; MEKAL~=~thermal plasma emission; GA~=~Gaussian emission line; WA~=~photoelectric absorption}
\label{tab1}
\begin{center}
\begin{tabular}{lccccccc} \hline \hline
Model & $N^{soft}_{\rm H}$ & $N^{hard}_{\rm H}$ & $\Gamma_{{\rm
hard}}$$^{\nabla}$ & $E_{{\rm break}}$ & $\Gamma_{{\rm soft}}$ & kT & $\chi^2/\nu$ \\
& ($10^{21}$~cm$^{-2}$) & ($10^{23}$~cm$^{-2}$) & & (keV) & & (keV) & \\ \hline
\multicolumn{8}{l}{XMM-Newton} \\
\#1 & $1.3 \pm^{1.2}_{1.0}$ & ... & $0.7 \pm^{0.3}_{0.4}$ & $2.3\pm^{0.8}_{0.4}$ & $2.1 \pm^{0.6}_{0.3}$ &... & 34.0/41 \\ 
\#2 & $1.2 \pm^{0.9}_{0.8}$ & $2.4 \pm^{1.0}_{0.8}$ & $2.0
\pm^{0.4}_{0.3}$ & ... & $\equiv \Gamma_{{\rm hard}}$ & ... & 31.1/41 \\
\#3a$^{\clubsuit}$ & $1.1 \pm 0.5$ & ... & $2.21
\pm^{0.19}_{0.14}$$^{\nabla}$ & ... & $\equiv \Gamma_{{\rm hard}}$ & ... & 34.8/42 \\
\#3b & $1.9 \pm ^{1.3}_{1.0}$ & ... & 2$^{\nabla,\ddag}$ & ... & $2.6
\pm 0.5$ & ... & 32.2/42 \\
\#4 & $0.9 \pm^{0.8}_{0.6}$ & ... & $2.2 \pm 0.4$$^{\nabla}$ & ... & ... & $1.3 \pm^{0.8}_{0.4}$ & 33.2/42 \\
\multicolumn{8}{l}{ASCA} \\
\#3a$^{\clubsuit}$  & $<0.8$ & ... &$1.2 \pm^{2.0}_{0.3}$$^{\nabla}$ &
... & $\equiv \Gamma_{{\rm hard}}$ & ... & 79.7/77 \\
\hline \hline
\end{tabular}
\end{center}

\noindent
$^{\nabla}$intrinsic, for the Compton-reflection model;
$^{\clubsuit}$``baseline model" in text; $^{\ddag}$fixed.

\end{table*}
The 
F-test for the addition
of emission line feature to the broken power-law
continuum is highly significant: $\Delta \chi^2/\Delta
\nu = 36.5/3$, which corresponds to a 99.9992\%
confidence level according to the F-test\footnote{Following 
Protassov et al. (2002) we have applied the F-test
without imposing any constraints on the
Gaussian profile normalization sign.}
A zoom of the spectral area around 6 keV is shown in Fig.~\ref{fig4}
(only pn data are shown for clarity), where a constant
linear rebinning in the energy space has been applied to
prevent weak, narrow-band features from being
visually smeared.
The emission feature clearly stands out against the underlying
continuum. We have simultaneously fitted the
non background-subtracted pn and MOS spectra in the 5.25-7.25~keV
energy band, using the C-statistics (which is appropriate for
spectra, whose channel photon distribution is purely Poissonian).
In this fit, the local pn (MOS)
background has been modeled with a simple power-law
(a good approximation in such a small energy range), with
normalization $1.41 \times 10^{-6}$
($2.61 \times 10^{-7}$)~keV~cm$^{-2}$~s$^{-1}$ and
photon index 0.38 (-~0.52). The local continuum has been
approximated by a power-law as well. A single narrow
emission line leaves residuals bluewards the energy
centroid. They can be accounted for either by a single
broad line, with centroid energy $E_{\rm c} = 6.44 \pm^{0.08}_{0.04}$~keV,
$\sigma = 100\pm^{100}_{50}$~eV and
intensity $I_{\rm c} =
(4.2 \pm^{1.8}_{0.8}) \times 10^{-6}$~cm$^{-2}$~s$^{-1}$,
or by a combination of three narrow lines, whose energies
are well consistent with neutral, He-like, and H-like iron.
(cf. Tab.~\ref{tab3}). Total net counts in
the line are $110 \pm 50$. Although these ``local fit" results
may be affected by uncertainties in the determination
of the local continuum, and need
therefore to be regarded with caution,
they represent a pathfinder to
\begin{table}
\caption{Properties of the emission line components for
a "local" (power-law and narrow Gaussian profiles
in the 5.25--7.25~keV energy band) and a "global fit"
(model \#3a in Tab.~\ref{tab1}). Errors are 1-$\sigma$
for one interesting parameter. EWs are calculated against
the proper continuum (details in text)}
\label{tab3}
\begin{tabular}{lcc} \hline \hline 
& "Local fit" & "Global fit" \\ \hline
$E^{(1)}_{\rm c}$~(keV) & $6.42 \pm^{0.04}_{0.02}$ & 6.4$^{\dag}$ \\
$I^{(1)}_{\rm c}$~($10^{-6}$~cm$^{-2}$~s$^{-1}$) & $3.4 \pm^{0.9}_{1.2}$ & $2.7 \pm^{0.7}_{0.8}$ \\
$EW^{(1)}$~(eV) & ... & 630 \\
$E^{(2)}_{\rm c}$~(keV) & $6.70 \pm^{0.12}_{0.18}$ & 6.7$^{\dag}$ \\
$I^{(2)}_{\rm c}$~($10^{-6}$~cm$^{-2}$~s$^{-1}$) & $0.9 \pm^{0.8}_{0.7}$  & $2.3 \pm^{0.9}_{1.4}$ \\
$EW^{(2)}$~(eV) & ... & 3400 \\
$E^{(3)}_{\rm c}$~(keV) & $6.97 \pm^{0.07}_{0.08}$ & 6.96$^{\dag}$ \\
$I^{(3)}_{\rm c}$~($10^{-6}$~cm$^{-2}$~s$^{-1}$) & $1.2 \pm^{0.9}_{0.7}$ & $<1.0$ \\
$EW^{(3)}$~(eV) & ... & $<1600$ \\ \hline \hline
\end{tabular}

\noindent
$^{\dag}$fixed

\end{table}
guide the global fitting.

The line complex has a very large Equivalent Width
($EW = 1.5 \pm^{0.5}_{0.4}$~keV). 
Equivalent widths of this order
can be most naturally explained if the continuum
underlying the emission line is partially or totally
obscured, whereas the line photons reach us through an optical
path, which does not intercept the obscuring matter (\cite{turner97},
\cite{matt00}).
In the broken power-law fit of the Mkn~668 spectra,
the photon index above $\simeq$2.3~keV
($\Gamma_{{\rm hard}} = 0.7 \pm 0.3$) is indeed
flatter than typically observed in unobscured
AGN (\cite{nandra97}, \cite{reeves00}), and therefore advocates
for obscuration of the nuclear emission. 

In light of the above results, we have tested a
scenario whereby the hard X-ray continuum is produced
by an heavily absorbed nuclear continuum,
whereas the soft excess and
the He- and H-like iron lines are produced by electron
scattering of the same nuclear
continuum. In practical terms, 
we have substituted the broken power-law continuum
with the combination of two power-laws,
photoabsorbed by column densities $N^{soft}_{\rm H}$
and $N^{hard}_{\rm H}$, respectively.
Even if the soft and hard photon indices
are tied together, the fit is good ($\chi^2 = 31.5/42$).
The flat high-energy spectrum is well represented
by a ``standard"
AGN spectral index ($\Gamma_{{\rm hard}} = 2.0\pm^{0.4}_{0.3}$)
absorbed by $N^{hard}_{\rm H} = 2.4\pm^{1.0}_{0.8} \times 10^{23}$~cm$^{-2}$.
In this scenario, however, the
equivalent width of the
neutral iron line ($EW \simeq 700$~eV) remains unexplained
(\cite{leahy93}). A comparatively good fit
($\chi^2 = 35.7/43$) is obtained if the hard X-ray emission
is modeled with a ``bare" Compton-reflection component
(model {\tt pexrav} in {\sc Xspec}; \cite{magdziarz95}),
therefore assuming that a Compton-thick absorber
($N_{\rm H} \ge \sigma_{\rm t}^{-1} \simeq 1.5 \times
10^{24}$~cm$^{-2}$),
totally suppresses the nuclear emission below
10~keV (\cite{matt99b}).
The large iron line EW ($\simeq 630$~eV) is naturally produced
in this scenario (\cite{ghisellini94}, \cite{krolik94}). 
Any transmitted component is constrained
to be absorbed by $N^{{\rm hard}}_{\rm H}
\approxgt 9 \times 10^{23}$~cm$^{-2}$ (for $\Gamma = 2.21$
and a warm scattering fraction of 5\%).
In Tab.~\ref{tab3} we summarize the properties of the
emission line complex in the ``global fit'' model,
if one assumes a 3-components
(neutral, He-like and H-like iron) decomposition.
The EW are calculated with respect to the
continuum against which they are supposedly produced:
the Compton-reflection continuum for the neutral component,
the scattered continuum for the ionized components.
The normalizations of the continuum components in this model are:
$N^{{\rm reflection}} = (1.6 \pm 0.5) \times 10^{-3}$, and
$N^{{\rm scattering}} =
(3.5 \pm^{0.5}_{0.4}) \times 10^{-5}$~keV~cm$^{-2}$~s$^{-1}$,
for the Compton-reflection and the scattering component, respectively.
They corresponds to observed fluxes
of $(0.6\pm^{0.6}_{0.2})$ and $(3.1\pm^{1.0}_{0.7}) \times
10^{-13}$~erg~cm$^{-1}$~s$^{-2}$ in the 0.5--2~keV
and 2--10~keV energy bands, respectively,
and to unabsorbed luminosities of
$(0.8\pm^{0.8}_{0.3})$ and $(4.1\pm^{1.3}_{0.9}) \times
10^{42}$~cm$^{-2}$ in the same energy bands.
We will refer to this model (\#3a in Tab.~\ref{tab2})
as ``baseline model" hereinafter.

An alternative origin for the ionized components of the
emission line complex is thermal emission in
an optically thin, collisionally ionized plasma
(\cite{boller03}). We have therefore tried to fit
the spectrum with a combination of a Compton-reflection
continuum, a single narrow Fe{\sc i} emission line, and
thermal emission component, using for the last
the {\tt mekal} implementation
in {\sc Xspec} (\cite{mewe85}). The fit is again
statistically acceptable. However, the temperature of the
thermal component ($kT \simeq 1.3$~keV, with $Z < 0.14 Z_{\odot}$)
is not high
enough to contribute significantly to the ionized lines emission
in the iron regime. The addition
of a further thermal component is not required from the
statistical point of view. It is also worth
mentioning that a model constituted by two
thermal components does not fit well the EPIC
spectra ($\chi^2/\nu$=~65.2/42).

\section{Comparison with previous X-ray observations}

\subsection{ASCA observations}

Mkn~668 was observed by ASCA between January 4 and January 6,
1999. Preliminary results on this observation are discussed by Guainazzi
et al. (2003), who stress that, due to the low statistics,
it is impossible to obtain a fully unambiguous spectral
deconvolution. We extracted spectra from the screened
event lists available at the HEASARC ASCA archive,
employing circular regions of
3$\arcmin$ radius around the source centroid, after
retaining only SIS events with standard (0, 2, 3, 4)
grades. Total exposure time was 65.5~ks (72.1~ks)
for the SIS (GIS) cameras. Background spectra were extracted from 
field-of-view regions, free from contaminating sources.
Response matrices
appropriate for the extracted spectra
were generated with {\sc Lheasoft} v5.1.
Applying the baseline model simultaneously  
to the spectra of all the ASCA instruments
one gets an acceptable fit ($\chi^2/\nu = 79.7/77$).
Both the continuum and the line best-fit
parameters are consistent with those derived from
the XMM-Newton observation, within the
admittedly large statistical
uncertainties. In particular,
a large
EW iron line is marginally detected in the ASCA observation
as well, suggesting, alongside with the flat hard X-ray
continuum, a Compton-reflection dominated spectrum.
The parameters of the iron line complex, when fit
with a single Gaussian profile, are: $E_{\rm c} = 6.7 \pm^{0.4}_{0.2}$~keV,
$\sigma = 250 \pm^{350}_{250}$~eV,
$I_{\rm c} = (7 \pm^8_4) \times 10^{-6}$~cm$^{-2}$~s$^{-1}$
(corresponding to an $EW = 850$~eV).
The 2--10~keV flux observed by ASCA is almost double
as measured during the XMM-Newton observation:
$(6.1\pm^{1.3}_{1.2}) \times 10^{-13}$~erg~cm$^{-2}$~s$^{-1}$.
This is suggestive of
an historical change of the nuclear emission
output power by a factor of 2 between the two observations,
although a contribution from a larger brightening of
one of the sources detected by XMM-Newton and encompassed
in the large ASCA aperture cannot be ruled out.
On the other hand, the 0.5--2~keV ASCA flux is
consistent with the later XMM-Newton
measurement:
$(0.9\pm^{0.4}_{0.3}) \times 10^{-13}$~erg~cm$^{-2}$~s$^{-1}$.

An earlier (July 17 1998) and shorter (GIS exposure time
33~ks) ASCA observation produced a 6$\sigma$ detection,
corresponding to a total 0.5--10~keV unabsorbed flux
of $(5 \pm 2) \times 10^{-13}$~erg~cm$^{-2}$~s$^{-1}$.
No useful spectral information can be derived from this
observation.

\subsection{ROSAT observations}

\begin{figure}
   \centering
   \includegraphics[width=8cm]{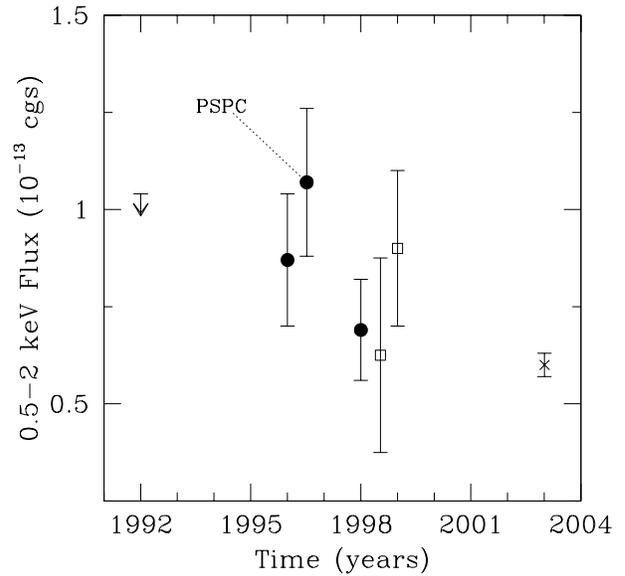}
      \caption{Historical 0.5--2~keV flux light curve of Mkn~668.
		Measurements are corrected for absorption.
		Error bars represent the propagation of the 1-$\sigma$
		uncertainties on the count rates. {\it Filled circles}
		and {\it upper limit}: ROSAT; {\it empty squares}:
		ASCA; {\it cross}: XMM-Newton
              }
         \label{fig8}
   \end{figure}

Mkn~668 was observed 4 times by ROSAT, one of them with
the Position Sensitive Proportional Counter
(PSPC) detector, and the remaining with the
High Resolution Imager (HRI). Preliminary
results of the former observation are
discussed by Zhang \& Marscher (1994). We have reanalyzed the
ROSAT public data, extracting calibrated event lists
from the public archive. The results of the analysis
are summarized in
Tab.~\ref{tab4}. The PSPC spectrum can be adequately fit
\begin{table}
\caption{Summary of spectral results for the ROSAT observations
of Mkn~668 (details in text).
The superscript refers to HRI (``H'') or PSPC
(``P'') observations}
\label{tab4}
\begin{tabular}{lccc} \hline \hline
Date & $N_{\rm H}$ & $\Gamma$ & 0.1-2.4~keV \\ 
& ($10^{21}$~cm$^{-2}$) & & rate ($10^{-3}$~s$^{-1}$) \\ \hline
9/1/92$^H$ & ... & ... & $< 2.4$ \\
23/1/96$^H$ & ... & ... & $2.0 \pm 0.4$ \\
10/7/96$^P$ & $0.8 \pm^{3.1}_{0.5}$ & $2.4 \pm^{2.1}_{1.1}$ & $7.9 \pm 1.3$ \\
12/1/1998$^H$ & ... & ... & $1.6 \pm 0.3$ \\ \hline \hline
\end{tabular}

\end{table}
in the 0.1--2.4~keV energy band by an absorbed power-law
($\chi^2/\nu = 17.6/20$).  In Fig.~\ref{fig8} we compile eleven years
of Mkn~668
soft X-ray flux measurements.
The HRI fluxes
have been extrapolated from the observed count rate, assuming
an absorbed power-law with $N_{\rm H} = 1.4 \times 10^{21}$~cm$^{-2}$
and $\Gamma = 2.2$ (best-fit
of the XMM-Newton EPIC spectra in the 0.5--2~keV
energy band alone). Caution must be
applied in interpreting this plot, due to the 
uncertainties induced by the unknown spectral shape in the ROSAT/HRI
measurements, and to the potential contamination of serendipitous
sources in the large ASCA aperture. Nonetheless, the
observed soft X-ray variability dynamical range in Mkn~668 is
less than a factor 50\%. The XMM-Newton observation has possibly
caught one of its faintest soft X-ray states ever.
 
\section{Discussion}

\subsection{Mkn~668: a Compton-thick Seyfert 1 galaxy}

\begin{figure}
   \centering
   \includegraphics[angle=-90,width=8cm]{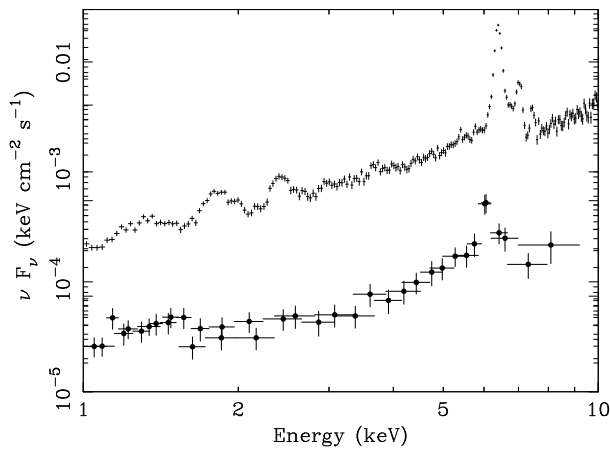}
      \caption{X-ray SEDs for the Circinus Galaxy ($z=0.0015$;
		{\it top}, pn data)
		and the Mkn~668 ($z=0.077$; {\it bottom}, EPIC data)
		nuclei
              }
         \label{fig5}
   \end{figure}

The X-ray spectrum of Mkn~668 exhibits two distinct components.
At energies $E \approxgt 2.5$~keV the spectrum is
remarkably flat. This
evidence, alongside with the prominent K$_{\alpha}$ fluorescent
emission line, represents a clear indication for a
Compton-reflection dominated spectrum (\cite{matt00}
and references therein). The direct view of
the primary nuclear emission is hindered by photoelectric absorption
with a column density $N_{\rm H} \approxgt 9 \times 10^{23}$cm$^{-2}$.
This is confirmed
by the overall similarity between the X-ray Spectral
Energy Distributions (SEDs) of Mkn~668 and of the Circinus Galaxy
(cf. Fig.~\ref{fig5}). The latter is the closest
Compton-thick AGN known; its flat hard X-ray continuum
and prominent line emission complex are interpreted in terms of
Compton-reflection 
of an otherwise invisible (below 10~keV) nuclear emission (\cite{matt99a},
\cite{molendi03}).

If there are little doubts that the X-ray appearance of
Mkn~668 is that of a highly obscured AGN, its optical
spectrum shows features more typical of a
Broad Line Radio Galaxy (\cite{blake70}).

Marziani et al. (1993) discuss the peculiar
properties of the broad optical lines in
Mkn~668. The peaks of the
broad H$_{\alpha}$ and H$_{\beta}$ components are
displaced to the red with respect to the centroid
of the narrow components of the same lines. The
observed correlation between line luminosity and
centroid displacement is well explained by radiative
acceleration in a system of outflowing clouds.
Marziani et al. (1993) 
therefore conclude that broad line emission in
this galaxy partly occurs in a cone of half-opening
angle $\simeq$12$^{\circ}$ seen at inclination angles
$ \imath \approxlt 30^{\circ}$.

To our knowledge, Mkn~668 is the first known non-transient
Seyfert~1 galaxy, whose X-ray spectrum is
Compton-reflection dominated, hence suggesting
that the nucleus is covered by a Compton-thick
absorber. The only analogous case reported in the
literature is the ``off-state'' of NGC~4051
(\cite{guainazzi98}).
The X-ray spectrum of this Narrow Line Seyfert~1
Galaxy was discovered by BeppoSAX
to be Compton-reflection dominated during a
$\simeq$100~days-long period when the total
X-ray output decreased by a factor of $\simeq 10^2$
with respect to its standard activity level
(\cite{uttley99}). 

However, no evidence exists in Mkn~668 for large
variations of the AGN X-ray output. In hard X-rays, only the
measurements described in this paper are available,
which limit the variability dynamical range above
2~keV within a modest factor of 2. In the soft
X-ray range, where a better temporal coverage is available,
there is no evidence for a variability dynamical range
larger than 50\% (cf. Fig.~\ref{fig8}).
There is therefore no direct evidence so far that extreme X-ray
variability is the culprit for the discrepancy
between the optical classification and the
X-ray spectral properties in Mkn~668.

\subsection{The soft X-ray spectrum}

The soft X-ray spectrum of Mkn~668 is dominated by a
steeper component, whose origin cannot be
unambiguously unveiled on the statistical basis alone.
No clear evidence for soft X-ray emission lines
is found. The 90\% upper limits on the equivalent
widths of mid-Z He-like K$_{\alpha}$ fluorescent lines
({\it i.e.}: {\sc Ne ix}; {\sc Mg xi}; {\sc Si xiii};
{\sc S xv}) are in range
40--80~eV (against the scattering continuum),
therefore a factor 3 to 10
lower than typically observed in
``warm scattered'' Seyfert~2 galaxies (\cite{guainazzi99}).
On the other hand, the profile of the emission complex
in the iron fluorescent regime, significantly
more complex than a single unresolved component,
suggests a contribution from highly ionized
iron species. The EW of the iron He-like emission
line ($EW = 3.4 \pm^{2.0}_{1.3}$~keV) is
consistent with optically thin resonant scattering
(\cite{matt96}) by plasma with standard solar
abundances. A substantial
contribution to the ionized iron line
from thermal plasma is unlikely.

Although the warm scattering scenario
nicely explains simultaneously the soft
X-ray continuum and the Fe{\sc xxv} fluorescent
emission,
the agreement between the
spectral energy indices in the radio
($\alpha_{{\rm radio}} \simeq -1.1$) and in
the soft X-ray ($\alpha_{\rm X} \simeq -1.2$)
may indicate that at least part of the soft X-ray
photons are produced by non-thermal processes,
occurring in the same electron population
responsible for the emission in the radio band.
However, if we
make use of Kameno et al. (2000)
best-fit synchrotron model for the radio spectrum of the
NE lobe,
assuming free-free absorption with $\tau_{{\rm ff}} = 6.4 \pm 0.4$,
the extrapolation underpredicts the soft X-ray counts by at least
0.5 dex (the discrepancy in the hard X-ray band is obviously larger).
A monitoring of the 4.9~GHz flux in
Mkn~668 between 1975 and 1995 detected only
a 20\% gradual {\it decrease} of the radio flux
(\cite{stanghellini97}), ruling out an explanation of the
discrepancy in terms of lobe radio flux variability.
There is actually evidence for a {\it steepening} of
the lobe spectrum with increasing frequency
(\cite{dallacasa00}), with $\alpha_{radio}$ decreasing
from $\simeq$-0.7 in the 8.4--15~GHz to $\simeq$-1.4
in the 15--22~GHz frequency range. Finally,
the radio spectra available so far
do not resolve the hotspot,
and represent therefore a blending between
the hotspot and the surrounding lobe.
Although only strictly
simultaneous and aperture-matched
observations of Mkn~668 in the radio and
X-ray bands will be able to definitely rule out this
possibility, there is no clear evidence so far that synchrotron
emission by an electron population consistent
with the 1--15~GHz spectrum
represents a major contribution to the
soft X-ray spectrum.

Alternatively, Inverse Compton (IC)
have been suggested to be the dominant process
in large-scale X-ray jets
(\cite{tavecchio00}, \cite{celotti01}),
particularly in GPS quasars (\cite{siemiginowska02},
\cite{siemiginowska03}). The seed photons scattered
off the lobe electrons may be
the same synchrotron photons responsible for
the radio hotspot emission (Synchrotron
Self-Compton, SSC, model),
Cosmic Microwave Background photons (CMB),
or could be locally produced in the AGN environment.
In the SSC scenario, the ratio between the
IC and the synchrotron luminosity is given
by the ratio between the photon ($u_{\rm e}$) and
the magnetic energy density ($u_{\rm m}$), which in
turn can be expressed as (\cite{tavecchio98}):
\begin{equation}
\frac{u_{\rm e}}{u_{\rm m}} \sim 6 \times 10^{-4} \nu_{{\rm 8.4}} T^5_{{\rm 11}}
\end{equation}
where $T_{{\rm 11}}$ is the brightness temperature
at the frequency $\nu_{{\rm 8.4}}$, in units of
8.4~GHz. At this frequency, $T_{{\rm 11}}=0.5$ for
Mkn~668 (\cite{ghisellini93}).
It is therefore implausible that this
process significantly contributes to the soft X-rays.

The energy
density of the CMB at Mkn~668 redshift
is $\sim 5 \times 10^{-13} \gamma$~erg~cm$^{-3}$,
where $\gamma$ is the electron Lorentz factor.
One may estimate the magnetic field in the
hotspot region $B_{\rm IC}$
applying the Harris \& Krawczynski (2002)
prescription to the observed ratio
between the radio ($S_{\rm 5 \ GHz} \simeq 2.7$~Jy;
\cite{stanghellini96}) and the soft X-ray
($L_{\rm X} = 8 \times 10^{41}$~erg~s$^{-1}$)
luminosity. Assuming an inclination angle
of 15$^{\circ}$ (45$^{\circ}$)\footnote{Stanghellini
et al. (1997) estimate $\imath \simeq 45^{\circ}$
on the basis of the Mkn~668 optical properties,
and assuming that the radio axis is perpendicular
to the plane of the galaxy}, one
gets $B_{\rm IC} \sim 3$~$\mu$G
(5~$\mu$G). This is far too low
with respect to the minimum energy condition
magnetic field ($B_{\rm me}$; \cite{pacholcyzk70}).
The observed 5~GHz integrated flux density
in Mkn~668, implies
$B_{\rm me} \sim 60 \  \theta_{{\rm sec}}^{-4/7}
( \sin \imath )^{-3/7}$~$\mu$G
(\cite{miley80}), where $\theta_{{\rm sec}}$
is the hotspot size in arcseconds. In Mkn~668
$\theta_{{\rm sec}} \simeq$0.7$\times$0.2~mas (\cite{barone03}).

The local AGN radiation energy density, $u_{\rm AGN}$
potentially available
for Compton scattering should be dominated by
the remarkably large Far Infrared Emission ($L_{\rm FIR} \simeq
1.4 \times 10^{11} L_{\odot}$; \cite{mazzarella91}).
An exact estimate of the energy density at the radio hotspots
would require a detailed knowledge of the geometry
of the circumnuclear region, which is largely unknown.
If the FIR-emitting dust is located within
the radio structure (see Sect.~4.3.1), and
the radio hotspots have a unity filling factor
we estimate $u_{\rm AGN} \sim 6 \times 10^{-12} (\Omega_{\rm dust}/2
\pi)$~erg~cm$^{-3}$, where $\Omega_{\rm dust}$ is
the solid angle subtended by the radio hotspots to
the dust. For moderate electron Lorentz factors,
the local AGN radiation energy density can indeed exceed the
CMB energy density at the hotspots, if
dust and radio lobes cover regions of comparable
size and/or are close.

\subsection{The nature of the X-ray absorbers in Mkn~668}

The spectral analysis of the XMM-Newton observation of the
Mkn~668 unveils the presence of two different absorbing
systems. Together with the Compton thick absorber
covering the direct line-of-sight to the nucleus, an
additional absorbing system with $N_{\rm H} \sim 10^{21}$~cm$^{-2}$
covering the soft X-ray spectrum is required by all models
presented in Sect.~2. In this Section we compare the
X-ray and radio results, and derive some
physical properties of the
absorbing systems.

\subsubsection{The Compton-thick absorber}

We first ask whether the Compton-thick matter could be
responsible for the free-free absorption of the radio
hotspots (\cite{kameno00}), and could therefore be located at a distance
from the active nucleus at least
larger than the half-separation
between the two lobes ($\simeq$5~pc).
An estimate of the AGN intrinsic power
in Mkn~668 can be inferred from the
O{\sc [iii]} flux ($5 \times 10^{-14}$~erg~cm$^{-2}$~s$^{-1}$;
\cite{marziani93}). According to the correlation between
the O{\sc [iii]} and the 2--10~keV luminosity
[$\log (L_{{\rm 2-10 \ keV}}/L_{{\rm O[III]}}) \sim 1.8$; \cite{mulchaey94},
\cite{maiolino98}], the intrinsic 2--10~keV AGN luminosity should be
$\sim 9 \times 10^{43}$~erg~s$^{-1}$. If the
size of the scattering medium is of the same order of
its distance from the active nucleus, one has:
\begin{equation}
l_{{\rm sc}} = \frac{\sigma_{\rm T} L_{{\rm int}}^2}{\xi L_{{\rm sc}}}
\biggl ( \frac{\Omega}{4 \pi} \biggr )
\end{equation}
where $l_{\rm sc}$ is the size of the scattering plasma, $L_{{\rm sc}}$ is the
{\it observed} 2--10~keV luminosity of the scattering component
($2 \times 10^{42}$~erg~s$^{-1}$), $n$ is the
scatterer particle density, $\xi \equiv L_{{\rm int}}/l_{\rm
sc}^2n$
is the ionization parameter, and
$\Omega/(4 \pi)$ is the solid angle subtended by the scattering
matter to the nucleus. If we express: $\xi \equiv 500 \xi_{500}$
(the detection of He-like iron fluorescent
emission line associated with the scattering continuum
implies a ionization parameter of at least a few hundreds);
and $\Omega/(4 \pi) \equiv 0.01 [\Omega/(4 \pi)]_{0.01}$
(corresponding to a cone with a half-opening angle of 15$^{\circ}$),
$l_{\rm sc} \sim \xi_{500}^{-1} \ [\Omega/(4 \pi)]_{0.01} \ 0.02$~pc.
This would be too a small size for the scattered continuum
to be visible beyond an homogeneous distribution
of Compton-thick matter covering the
radio hotspots and the nucleus. This discrepancy
is not solved even if $\Omega/(4 \pi) = 0.25$
is assumed, believed to be a reasonable value in the
archetypical Compton-thick Seyfert~2 galaxy NGC~1068 (\cite{evans91}).
We conclude therefore that the Compton-thick matter 
must be located within
the separation between the radio hotspots, in agreement to
typical estimates of the torus size in Seyfert galaxies
(\cite{greenhill96},
\cite{greenhill97}, \cite{guainazzi00}, \cite{bianchi01}).

From its FIR luminosity, Knapp et al. (1990) estimate that
$\sim 5 \times 10^8 M_{\odot}$ of dust must be present in
the core of Mkn~668. If this dust is in the form of a
torus within 10 parsecs from the
core, the torus column density
must exceed $3 \times 10^{25}$~cm$^{-2}$ (in agreement with
the X-ray results). The
physical conditions in the torus can be investigated,
using the stability conditions for X-ray illuminated torii
discussed by Neufeld et al. (1994) and Maloney (1996).
The condition for a torus to be
fully molecular if that $\log (\xi_{{\rm eff}}) < -3.4$,
where $\xi_{{\rm eff}}$ is the ``effective'' ionization
parameter at the outer side of the torus, taking into
account self-shielding of the ionizing continuum. Following
Maloney (1996), $\xi_{{\rm eff}}$ can be expressed as:
\begin{equation}
\xi_{{\rm eff}} = 0.17 L_{{\rm 44}} N_{{\rm 24}}^{-0.9} n_{\rm 6}^{-1}
r_{{\rm pc}}^{-2}.
\end{equation}
$L_{{\rm 44}}$ is the X-ray luminosity in units of
$10^{44}$~erg~s$^{-1}$, $N_{{\rm 24}}$ is the column density
in units of $10^{24}$~cm$^{-2}$, $n_{\rm 6}$ is the
particle density in units of $10^6$~cm$^{-3}$,
and $r_{\rm pc}$ is the distance between the torus and
the active nucleus in parsecs. Moreover, a torus
is in steady state if its pressure 
exceeds the radiation pressure. If only the gas pressure is relevant:
\begin{equation}
n_{\rm 6} r_{{\rm pc}}^2 > 10^2 L_{{\rm 44}} T_{\rm 3}^{-1}
\label{eqn1}
\end{equation}
where $T_{\rm 3}$ is the average torus temperature
(600~K for a molecular torus: \cite{maloney96}).
From Eqn.\ref{eqn1}: $\log n_{\rm 6} + 2 \log r_{{\rm pc}} > 4.15$,
which implies $\log \xi_{{\rm eff}} < -4.75$,
in agreement with the molecular hypothesis.

\subsubsection{The Compton-thin absorber}

If the jet is opening its way through the ambient
interstellar medium, each radio lobe is surrounded by a
bow shock, whose precursor highly ionized clouds may
give rise to free-free absorption of the radio emission.
Philstr\"om et al. (2003) estimate
from ram pressure arguments a particle
density of $\simeq$30~cm$^{-3}$ at the external
surface of the bow shock generated by
the ``drilling jet'' (\cite{bicknell97}).
If the absorbing matter
is distributed in a shell surrounding the
radio hotspots, its radial thickness $\Delta r_{\rm ab}$ is
\begin{equation}
\Delta r_{\rm ab} = \frac{N_{\rm H}}{n_{\rm H}} \simeq 10~pc
\end{equation}
This implies that the radio and X-ray (Compton-thin) absorbing
media could coincide in a
layer preceding the ``drilling'' young
radio jet. Such a medium would not
contribute significantly to the X-ray emission, even
if heated up to X-ray emitting temperatures.
The expected 0.1--10~keV isotropic luminosity of a 1~keV
plasma in such a geometry would be at most
$\sim 2 \times 10^{37}$~erg~s$^{-1}$.

The measurement of hotspot recession velocities
suggests that Mkn~668 hosts a jet
still
in the infancy of its development (\cite{stanghellini02},
\cite{lister03}). 
Interestingly enough, however, the
post-shock particle density exceeds the average value
($n \sim 1$~cm$^{-3}$), sufficient to
ensure
the confinement of the radio jet on scales $\approxlt 1$~kpc
(\cite{deyoung93}). The ultimate fate of the jet
in Mkn~668 may therefore not have been written yet.

\begin{figure}
   \centering
   \includegraphics[width=8cm]{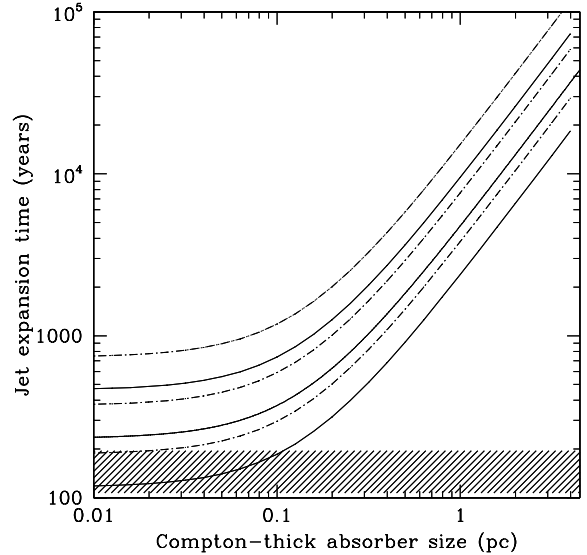}
      \caption{Jet expansion time as a function of the physical
		thickness of the Compton-thick absorber.
		{\it Solid lines} correspond to
		$\imath = 15^{\circ}$, {\it dashed-dotted
		lines} to $\imath = 45^{\circ}$. From
		{\it top} to {\it bottom}, curves plotted
		in the same style correspond to values
		of the jet opening angle of
		20$^{\circ}$, 10$^{\circ}$, and 5$^{\circ}$,
		respectively. The {\it light shaded area}
		corresponds to the dynamical age estimate
		from the hotspot expansion velocity.
              }
         \label{fig11}
   \end{figure}

In the light of the above results, we can revise the
issue of the parsec-scale radio activity age in
Mkn~668. Carvalho (1998) demonstrated that the
presence of a dense medium in the core of the host
galaxy can significantly delay the propagation
of the jet, leading to grossly inaccurate estimates
of the evolution time scales.
If the jet propagates under ram pressure equilibrium
through a homogeneous external medium, its expansion
velocity $v_{\rm j}$ can de derived from (\cite{carvalho98})
\begin{equation}
m_{\rm H} n_{\rm e} v^2_{\rm j} \simeq \frac{1}{3} u
\end{equation}
where $n_{\rm e}$ is the electron density and
$u$ the jet total internal energy density. The expansion
time to a distance $R_0$ can be therefore expressed
as (\cite{scheuer74}, \cite{carvalho85}):
\begin{equation}
t_{\rm e} \sim 4 \times 10^5 R_0^{1.5} (\cos \imath)^{-1.5} L_{\rm
inj}^{-0.5} \Omega^{0.5} N^{0.5}_{24} \ s
\end{equation}
where $L_{\rm inj}$ in the luminosity injected in the
jet,
and $\Omega$ is the solid angle subtended by
the jet at $R_0$. We assume for Mkn~668
$R_0 = 5$~pc, and
$L_{\rm inj} \simeq L_{\rm kin} \sim 2 \times 10^{44}$~erg~s$^{-1}$
(\cite{celotti97}). In Fig.~\ref{fig11} we show
$t_{\rm e}$ as a function of the physical size of the
Compton-thick absorber, in a simple radial geometry where
a fraction $f_{\rm thick} R_0$ of the optical path along
the jet is (actually was) occupied
by homogeneously distributed matter
with number density
corresponding to a total
column density $5 \times 10^{25}$~cm$^{-2}$,
and the remaining fraction (1-$f_{\rm thick}$)$R_0$
by matter corresponding to a total column density
$10^{21}$~cm$^{-2}$. Large inclination
angles, or large thickness of the Compton-thick
absorber even for moderate inclinations can significantly
brake the jet, leading to an underestimate
of its evolution time by one-two orders of
magnitudes.

\subsection{Are the AGN in GPS galaxies obscured?}

\begin{figure*}
   \centering
\hbox{
   \includegraphics[width=8cm]{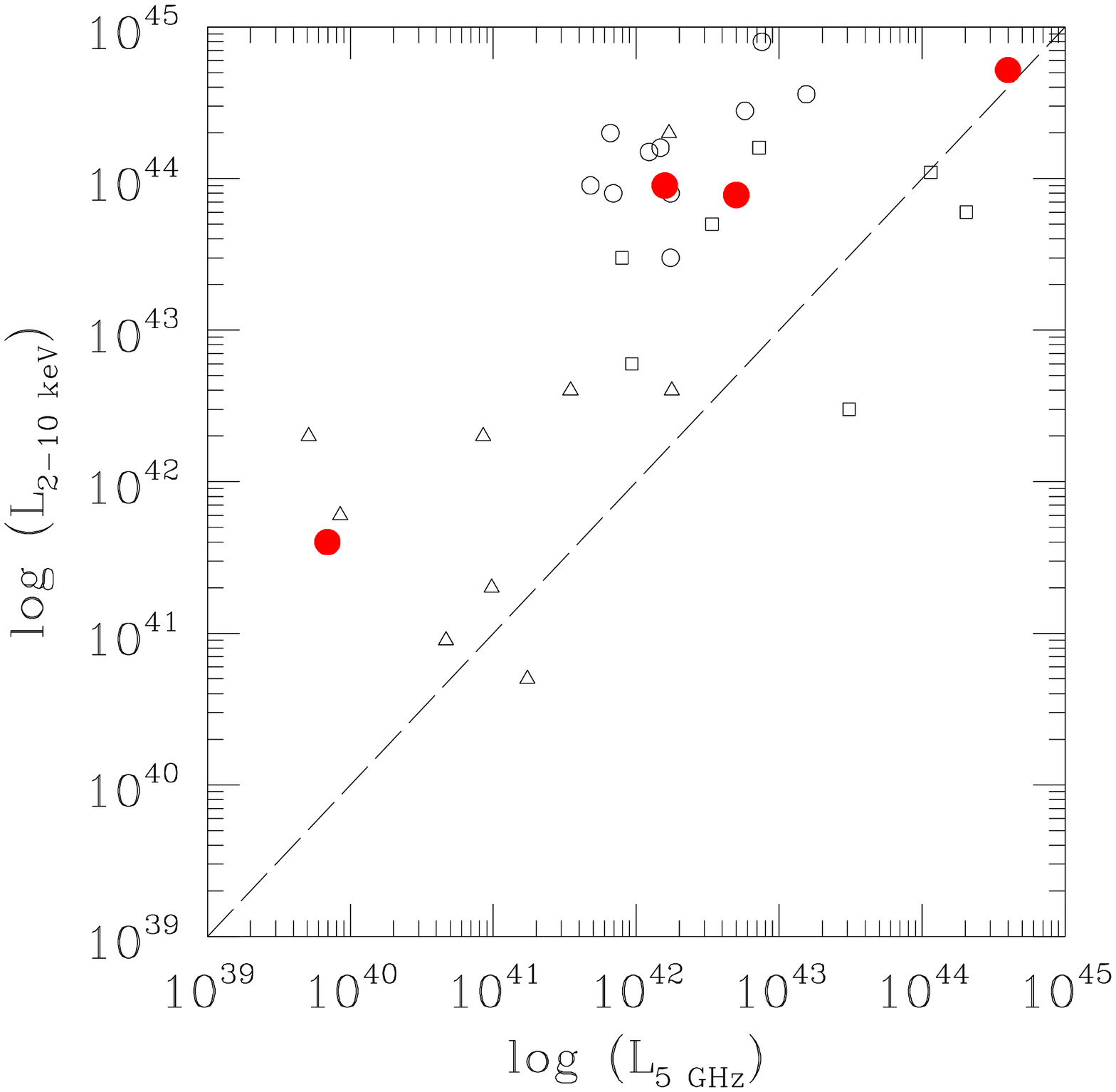}
\hspace{0.5cm}
   \includegraphics[width=8cm]{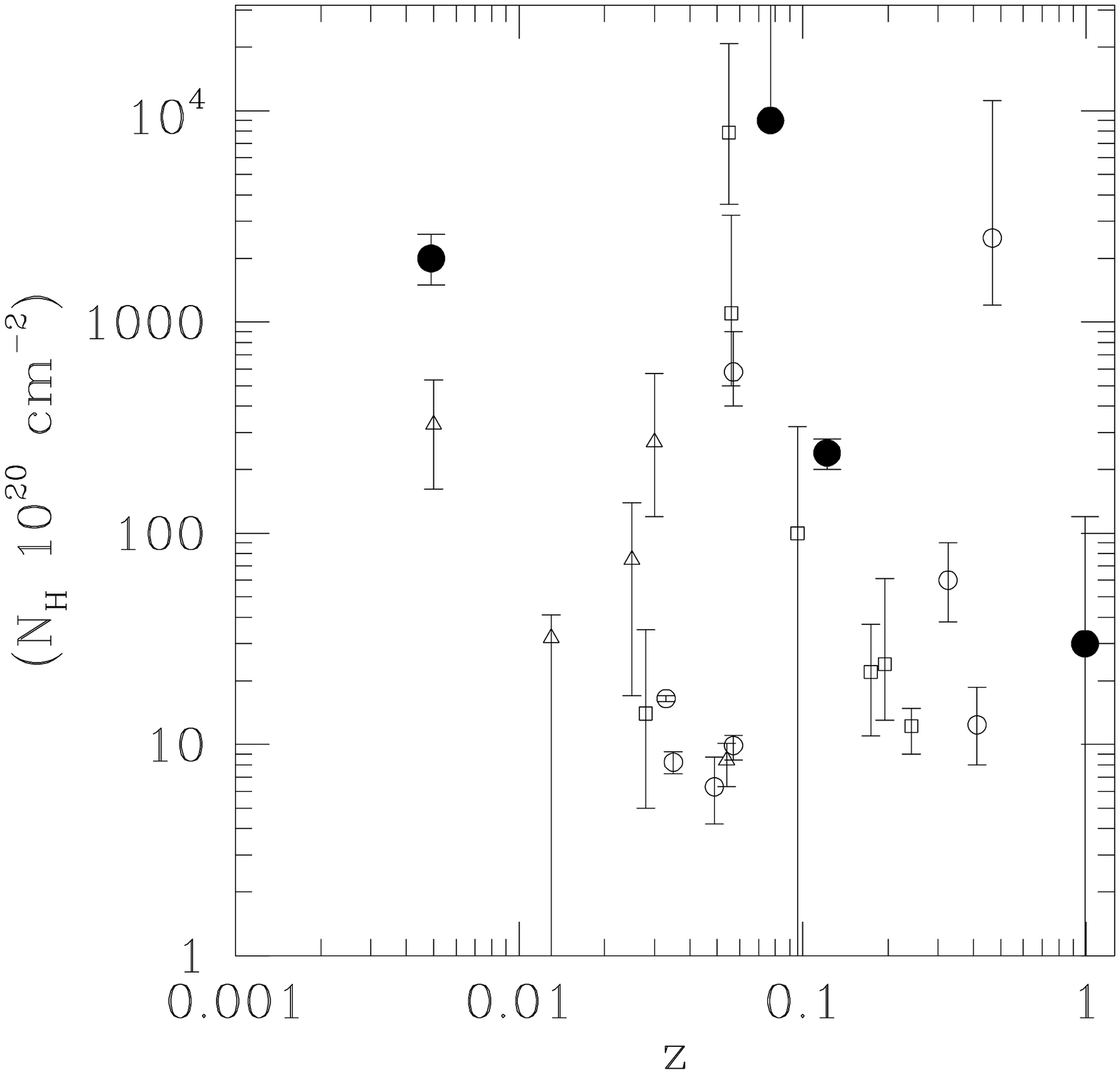}
}
      \caption{{\it Left panel}: 5~GHz versus 2--10~keV
		luminosity for the
		radio galaxies of the Sambruna et al. (1999)
		ASCA sample, and the GPS galaxies for which
		hard X-ray measurements are available.
		{\it Empty circles}: broad-line radio galaxies;
		{\it empty squares}: narrow-line radio galaxies;
		{\it empty triangles}: radio galaxies;
		{\it filled circles}: GPS galaxies.
		{\it Right panel}: $N_{\rm H}$ versus redshift distribution
		for the same sample.
              }
         \label{fig6}
   \end{figure*}

In the left panel of Fig.~\ref{fig6} we show the 2--10~keV
luminosity versus the 5~GHz luminosity for the radio-galaxy
sample of Sambruna et al. (1999)
and for the GPS galaxies, for which hard X-ray measurements
are available so far\footnote{NGC~1052, Guainazzi et al. 2000; 1345+125,
O'Dea et al. 2000; Mkn~668, this paper; Q2127+040. Siemiginowska
et al., in preparation}. The GPS galaxies - which span a range
of 5 decades in both radio and X-ray luminosity - do not show
any deviations from the behavior of ``standard'' radio
galaxies. On the other hand, 
Fig.~\ref{fig6} shows the dependence of the X-ray column density
on redshift\footnote{Whenever more than one absorbing system is
detected, the plot in Fig.~\ref{fig6} displays that with the largest
column density}. 3 out of 4 GPS galaxies exhibit an
absorbing column density larger than $10^{22}$~cm$^{-2}$,
whereas only $\simeq$30\% the Sambruna et al. radio galaxies do.
Drawing firm conclusions on this issue from the 
currently available data
is still impossible.
However, they
are at least
consistent with GPS galaxies hosting AGN covered on the average by
larger X-ray obscuration than large-scale radio galaxies.
The ongoing XMM-Newton program
will allow us to enlarge the sample of GPS galaxies with high-sensitivity
hard X-ray spectroscopic measurements,
and may therefore provide the final answer to this
unsettled issue

\section{Conclusions}

Despite of its optical appearance as a Seyfert~1 galaxy, the
X-ray spectrum of the GPS galaxy Mkn~668
exhibits typical features of obscured AGN:

\begin{itemize}

\item above $\simeq$2.5~keV, the spectrum is flat and
characterized by a prominent
($EW \simeq 600$~eV) neutral iron K$_{\alpha}$ emission line.
This suggests that Compton-reflection dominates
in this energy range, alongside
with fluorescence above a heavily absorbed primary nuclear continuum.

\item
the column density of the absorber covering the primary
emission
is constrained to be larger than
$9 \times 10^{23}$~cm$^{-2}$; however values larger
than several $10^{25}$~cm$^{-2}$ are more likely
if the absorber
reprocesses the AGN emission into the remarkably intense
observed FIR luminosity

\item the bulk of the soft X-rays
is due to a steeper ($\Gamma \simeq 2.0-2.5$)
power-law, absorbed by a further column density $N_{\rm H}^{{\rm soft}}
\sim 10^{21}$~cm$^{-2}$. We associate this component with
the observed fluorescent K$_{\alpha}$ emission from He-like iron.

\item (warm)-scattering of the nuclear
radiation is the most
straightforward explanation for the soft X-ray emission
and the Fe{\sc xxv} fluorescent K$_{\alpha}$ iron
line. Nonetheless, a contribution to soft
X-rays from IR photons up-scattered
by the hotspot electron population cannot be ruled out,
although a quantitative estimate would require
a detailed knowledge of the
galaxy core geometry.

\item we locate the Compton-thick absorber
in the innermost region of the galaxy core, within
the radio hotspots.
In order to be visible, the scattering plasma
needs to be located beyond the visible outer rim of the
Compton-thick absorbing system
(what is commonly observed in Seyfert 2 galaxies), probably
colocated or at least along the same line-of-sight as the
unobscured BLR
(what is instead {\it not} usual in Seyfert~2 galaxies,
where BLRs are hidden from out direct view)

\item we identify the X-ray Compton-thin absorber with matter
responsible for free-free absorption of the radio
source. In this scenario, this matter is probably distributed in
a $\simeq$10~pc shell
of relatively dense ($n \simeq 30$~cm$^{-3}$) matter
compressed by the jet bow shock.

\end{itemize}

The discrepancy between the optical classification and
the X-ray spectral properties in this object does
not fit the 0-th order unification scenario
for AGN (\cite{antonucci93}, \cite{barthel89}).
This might be
due to an uncorrelated
variability between the two bands, driven by large amplitude
variation in the overall energy output of the AGN,
as already observed in NGC~4051 (\cite{guainazzi98}),
or in the so-called ``changing-look'' Seyfert~2 galaxies
(\cite{matt03b} and references therein). However, there
is no evidence for large variation of the AGN power
in Mkn~668.
Another possibility is that the BLRs are seen
in reflected, rather than in direct, light
(see, {\it e.g.}, the discussion in Matt et al. 2003a).
Given the large EW of the broad optical lines
(\cite{marziani93}), this would imply a large
covering factor for the reflector. 
However, the discrepancy may be otherwise related
to the peculiar properties of this class of
radio object. ``Young'' AGN,
triggered by recent merging episodes,
may have more compact torii, which do not effectively cover
the direct view of the broad line region. Alternatively,
we may be observing the AGN through the line-of-sight
recently opened by the ``drilling'' jet, which frees
the view of the broad line regions. In this case,
a geometry such as a warped disk
may still ensure simultaneous obscuration
of the AGN. It is in principle possible as well,
that Mkn~668 hosts an ``intrinsic" type 2 AGN,
and that the BLRs are produced by the interaction
of gas with the expanding radio lobes. Bicknell et al. (1997)
make specific
predictions on the luminosity of [O{\sc iii}] and
H$_{\alpha}$+[N{\sc ii]} optical lines 
produced in this scenario.
The observed luminosities in Mkn~668
($\sim$~a few~$10^{40}$~erg~s$^{-1}$; Marziani et al. 1993)
are in good agreement with their model for
reasonable values of the ratio between the lobe energy
and the radio power (1~Jy at 1.7~GHz; Stanghellini
et al. 1997). However, the optical
lines are probably too broad to be produced by
such a mechanism. The upper limit on the velocity
of clouds accelerated by a bow shock can be estimated
by imposing that the shocked gas cooling time is
lower than the radio source dynamical age (\cite{odea02}):
$$
v_{cl} \approxlt 50 R_{pc}^{1/3} n^{1/3}_{ps,100}
v_{bs}^{-1/3} \le 60 \ \hbox{km~s$^{-1}$}
$$
where $R_{pc}$ is the distance (in parsecs)
from the nucleus to the bow shock, $n_{ps,100}$
is the post-shock density in units of 100~cm$^{-3}$,
and $v_{bs}$ is the bow shock velocity in units of
the speed of light ($\simeq$0.1 for Mkn~668; Stanghellini
et al. 2002).
On the other hand,
the observed H$_{\alpha}$ FWHM is $\simeq$6000~km~s$^{-1}$
(\cite{eracleous94}).
Moreover, a broad (FWHM~$\simeq$1450~km~s$^{-1}$)
H$_{\alpha}$ component was detected in polarized flux
(\cite{corbett98}). The
interstellar medium should be
distributed according to a rather peculiar geometry,
for bow shock
interactions to produce a broad polarized line.

An interesting suggestion was recently put forward
by Bellamy et al. (2003), who discovered
broad Pa${\alpha}$ line emission and a reddened
continuum in the young radio galaxy PKS~1549-79.
The compactness of the radio
emission in this object, together with the one-side
jet morphology, indicate that the radio axis is probably
aligned close to the line-of-sight. According to
the 0-th order AGN unification scenarios, the
BLR and the AGN in this object should be unobscured,
contrary to evidence. Bellamy et al. propose a scenario,
whereby PKS~1549-79 is an AGN in the early stages
of its evolution. The high obscuration
should be a transitory
phase, which will end as gas and dust are dissipated by
outflowing gas. The analogies between PKS~1549-79 and
Mkn~668 are remarkable, both exhibiting
simultaneously an obscured nucleus,
an outflowing BLR, and a small-scale radio jet.
However, the radio morphology of Mkn~668
exhibits clearly two radio-lobes, although their
power differs by about one order
of magnitude (\cite{kameno00}). This prevents
the line-of-sight to the
AGN in Mkn~668 from being too close to the radio
axis. Moreover, the detection of
a large-scale ($\simeq$30~kpc; \cite{debruyn90}) radio halo
may indicate that Mkn~668 could be undergoing
one of several phases of recurrent activity.

\begin{acknowledgements}
This paper is based on observations obtained with XMM-Newton, an ESA
science mission with instruments and contributions directly funded by
ESA Member States and the USA (NASA). This research has made use of
data obtained through the High Energy Astrophysics Science Archive
Research Center Online Service, provided by the NASA/Goddard Space
Flight Center and of the NASA/IPAC Extragalactic Database (NED) which
is operated by the Jet Propulsion Laboratory, California Institute of
Technology, under contract with the National Aeronautics and Space
Administration. MG gratefully acknowledges a period of visiting
astronomer at the Harvard-Smithsonian Center for Astrophysics,
supported by NASA Grant \#NAS8-39073,
where most of the research presented in this
paper was conducted. Useful discussions with F.Govino are acknowledged.
\end{acknowledgements}


\begin{thebibliography}{}

\bibitem[Antonucci 1993]{antonucci93} Antonucci R., 1993, ARA\&A, 31,
473

\bibitem[Baker et al. 1995]{baker95} Bekar J.C., Hunstead R.W.,
Brinkmann W., 1995, MNRAS, 277, 553

\bibitem[Barone 2003]{barone03} Barone K., 2003, Un. degree thesis

\bibitem[Barthel 1989]{barthel89} Barthel P.D., 1989, ApJ, 336, 606

\bibitem[Bellamy et al. 2003]{bellamy03} Bellamy M.J., Tadhunter C.N., 
Morganti R., et al., 2003, MNRAS, 344, L80

\bibitem[Bianchi et al. 2001]{bianchi01} Bianchi S., Matt G., Iwasawa
K., 2001, MNRAS, 322, 699

\bibitem[Bicknell et al. 1997]{bicknell97} Bicknell G.V., Dopita M.A., 
O'Dea C.P.O., 1997, ApJ, 485, 112

\bibitem[Blake et al. 1970]{blake70} Blake G.M., Argue A.N.,
Kenworthy C.M., 1970, ApJL, 6, 167

\bibitem[Boller et al. 2003]{boller03} Boller T., Keil R., Hasinger G., et al., 2003, A\&A, 411, 63

\bibitem[Carvalho 1985]{carvalho85} Carvalho J.C., 1985, MNRAS, 215,
463

\bibitem[Carvalho 1998]{carvalho98} Carvalho J.C., 1998, 329, 845

\bibitem[Celotti et al. 2001]{celotti01} Celotti A., Ghisellini G.,
Chiaberge M., 2001, MNRAS, 321, L1

\bibitem[Celotti et al. 1997]{celotti97} Celotti A., Padovani P.,
Ghisellini G., 1997, MNRAS, 286, 415

\bibitem[Corbett et al. 1998]{corbett98} Corbett E.A., Robinson A., Axon D.J., Young S., Hough J.H., 1998, MNRAS, 296, 721

\bibitem[Dallacasa et al. 2000]{dallacasa00} Dallacasa D., Stanghellini C., Centonza M., Fanti R., 2000, A\&A, 363, 887

\bibitem[de Bruyn 1990]{debruyn90} de Bruyn A.G., 1990, in ``Lecture
Notes in Physics 377, variability of Active Galaxies'', Duschl W.J.,
Wagner S.J. \& Camezind M. eds (Heidelberg), 105

\bibitem[de Vries et al. 1998]{devries98} de Vries W.H., O'Dea C.P.,
Baum S.A., et al., 1998, ApJ, 503, 156

\bibitem[De Young 1993]{deyoung93} De Young D.S., 1993, ApJ, 402, 95

\bibitem[Dickey \& Lockman 1990]{dickey90} Dickey J.M., Lockman F.J., 1990, ARA\&A 28, 215

\bibitem[Elvis et al. 1994]{elvis94} Elvis M., Fiore F., Mathur S.,
Wilkes B.J., 1994, ApJ, 425, 103

\bibitem[Eracleous \& Halpern 1994]{eracleous94} Eracleous M., Halpern J.P., 1994, ApJS, 90, 1

\bibitem[Evans et al. 1991]{evans91} Evans I.N., Ford H.C., Kinney
A.L., et al., 1991, ApJ, 369, L27

\bibitem[Fanti et al. 1995]{fanti95} Fanti C., Fanti R., Dallacasa D., 
et al., 1995, A\&A, 302, 317

\bibitem[Ghisellini et al. 1993]{ghisellini93} Ghisellini G., Padovani 
P., Celotti A., Maraschi L., 1993, ApJ, 407, 65

\bibitem[Ghisellini et al. 1994]{ghisellini94} Ghisellini G., Haardt
F., Matt G., 1994, MNRAS, 267, 743

\bibitem[Greenhill et al. 1997]{greenhill97} Greenhill L.J., Ellingsen 
S.P., Norris R.P., et al., 1997, ApJ, 474, L103

\bibitem[Greenhill et al. 1996]{greenhill96} Greenhill L.J., Gwinn
C.R., Antonucci R., Barvainis R., 1996, ApJ, 472 L21

\bibitem[Guainazzi et al. 1999]{guainazzi99} Guainazzi M., Matt G., Antonelli L.A., et al., 1999, MNRAS, 310, 10

\bibitem[Guainazzi et al. 1998]{guainazzi98} Guainazzi M., Nicastro
F.,  Fiore F., et al., 1998, MNRAS, 301, L1

\bibitem[Guainazzi et al. 2000]{guainazzi00} Guainazzi M., Oosterbroek T., Antonelli L.A., Matt G., 2000, A\&A, 364, L80

\bibitem[Guainazzi et al. 2003]{guainazzi03} Guainazzi M., Stanghellini C., Grandi P., 2003, MPE Report, 281, 261

\bibitem[Harris \& Krawczynski 2002]{harris02} Harris D.E.,
Krawczynski H., 2002, ApJ, 565, 244

\bibitem[Jansen et al. 2001]{jansen01} Jansen F., Lumb D., Altieri B., et al., 2001, A\&A 365, L1

\bibitem[Kameno et al. 2000]{kameno00} Kameno S., Horiuchi S., Shen
Z-Q., et al., 2000. PASJ, 52, 209

\bibitem[Knapp et al. 1990]{knapp90} Knapp G.R., Bies W.E., van Gorkom 
J.H., 1990, AJ, 99, 476

\bibitem[Krolik et al. 1994]{krolik94} Krolik J.H., Madau P., ${\rm
{\dot Z}}$ycky P., 1994, ApJ, 420, L57

\bibitem[Leahy \& Creighton 1993]{leahy93} Leahy D.A., Creighton J., 1993, MNRAS 263, 314

\bibitem[Lister 2003]{lister03} Lister M.L., 2003, ASP Conference
Series, in press (astroph/0301332)

\bibitem[Loiseau 2003]{loiseau03} Loiseau N., 2003, ``User's Guide to the XMM-Newton Science Analysis System, Issue 2.1, (XMM-Newton Science Operation Center:Villafranca del Castillo)

\bibitem[Magdziarz \& Zdziarski 1995]{magdziarz95} Magdziarz P.\& Zdziarski A.A., 1995, MNRAS 273, 837

\bibitem[Maiolino et al. 1998]{maiolino98} Maiolino R., Salvati M.,
Bassani L., et al., 1998, 338, 781

\bibitem[Maloney 1996]{maloney96} Maloney P.R., 1996, in ``Cygnus A -
Study of a Radio Galaxy'', Carilli C.L. \& Harris D.E. eds. (Cambridge 
University Press:Cambridge), 60

\bibitem[Marziani et al. 1993]{marziani93} Marziani P., Sulentic J.W., 
Calvani M., et al., 1993, ApJ, 410, 56

\bibitem[Matt et al. 2003a]{matt03a} Matt G., Bianchi S., Guainazzi M., et al., 2003, A\&A, 399, 519

\bibitem[Matt et al. 1996]{matt96} Matt G., Brandt W.N., Fabian A.C.,
1996, MNRAS, 280, 823

\bibitem[Matt et al. 2000]{matt00} Matt G., Fabian A.C., Guainazzi M., et al., 2000, MNRAS 318, 173

\bibitem[Matt et al. 1999a]{matt99a} Matt G., Guainazzi M., Maiolino R., 
et al., 1999, A\&A, 341, L27

\bibitem[Matt et al. 2003b]{matt03b}  Matt G., Guainazzi M., Maiolino
R., 2003, MNRAS, 342, 422

\bibitem[Matt et al. 1999b]{matt99b} Matt G., Pompilio F., La Franca
F., 1999, NewA, 4, 191

\bibitem[Mazzarella et al. 1991]{mazzarella91} Mazzarella J.M., Bothum 
G.D., Boroson T.A., 1991, AJ, 101, 2034

\bibitem[Mewe et al. 1985]{mewe85} Mewe R., Gronenschild E.H.B.M., van der Oord G.H.J., 1985, A\&AS, 62, 197

\bibitem[Miley 1980]{miley80} Miley G., 1980, AR\&AA, 18, 165

\bibitem[Molendi et al. 2003]{molendi03} Molendi S., Bianchi S., Matt G., 2003, MNRAS, 343, L1

\bibitem[Mulchaey et al. 1994]{mulchaey94} Mulchaey J.S., Koratkar A., 
Ward M.J., et al., 1994, ApJ, 436, 586

\bibitem[Nandra et al. 1997]{nandra97} Nandra K., George I.M., Mushotzky R.F., Turner T.J., Yaqoob T., 1997, ApJ 467, 70

\bibitem[Neufeld et al. 1994]{neufeld94} Neufeld D.A., Maloney P.R.,
Conger S., 1994, ApJ, 436, L127

\bibitem[O'Dea 1998]{odea98} O'Dea C., 1998, PASP, 110, 493

\bibitem[O'Dea et al. 2000]{odea00} O'Dea C., de Vries W.H., Worrall D.M., Baum S., Koekemoer A., 2000, AJ, 119, 478

\bibitem[O'Dea et al. 2002]{odea02} O'Dea C., de Vries W.H., Koekemoer A.M., et al., 2002, AJ, 123, 2333
 
\bibitem[Pacholcyzk 1970]{pacholcyzk70} Pacholcyzk A.G., 1970, ``Radio 
Astrophysics'' (San Francisco:Freeman)

\bibitem[Phillips \& Mutel 1982]{phillips82} Phillips R.B., Mutel R.L.,
1982, A\&A, 106, 21

\bibitem[Pihlstr\"om et al. 2003]{pihlstrom03} Pihlstr\"om Y.M.,
Conway J.E., Vermeulen R.C., 2003, A\&A, 404, 871
 
\bibitem[Polatidis \& Conway 2003]{polatidis03} Polatidis A.G., Conway 
J.E., 2003, PASA, 20, 69

\bibitem[Protassov et al. 2002]{protassov02} Protassov R., van Dyk
D.A., Connors A., Kashyap V.L., Siemiginowska A., 2002, ApJ, 571, 545

\bibitem[Readhead et al. 1996]{readhead96} Readhead A.C.S., Taylor
G.B., Xu W., et al., 1996, ApJ, 460, 612

\bibitem[Reeves \& Turner 2000]{reeves00} Reeves J.N., Turner M.J.L.,
2000, MNRAS, 316, 234

\bibitem[Sambruna et al. 1999]{sambruna99} Sambruna R., Eracleous M., Mushotzky R., 1999, ApJ, 526, 60

\bibitem[Sambruna et al. 2001]{sambruna01} Sambruna R., Netzer H., Kaspi S., et al., ApJ, 2001, 546, L13

\bibitem[Scheuer 1974]{scheuer74} Scheuer P.A.G., 1974, MNRAS, 166,
513

\bibitem[Siemiginowska et al. 2002]{siemiginowska02} Siemiginowska A., 
Bechtold J., Aldcroft T.L., et al., ApJ, 570, 543

\bibitem[Siemiginowska et al. 2003]{siemiginowska03} Siemiginowska A.,
Stanghellini C., Brunetti G., et al., 2003, ApJ, 595, 643
 
\bibitem[Snellen et al. 2002]{snellen02} Snellen I.A.G., Lehnert M.D., 
Bremer M.N., Schilizzi R.T., 2002, MNRAS, 337, 981

\bibitem[Stanghellini et al. 1997]{stanghellini97} Stanghellini C.,
Bondi M., Dallacasa D., et al., 1997, A\&A, 318, 376

\bibitem[Stanghellini et al. 1996]{stanghellini96} Stanghellini C.,
Dallacasa D., O'Dea C., et al., 1996, in ``Proceedings Second Workshop 
on GigaHertz Peaked-Spectrum and Compact Steep-Spectrum Sources'',
Snellen I.A., Schilizzi R.T., R\"ottgering H.J.A., Bremer M.N. eds.
(Leiden:Leiden Obs.). 4

\bibitem[Stanghellini et al. 2002]{stanghellini02} Stanghellini C.,
Liu X., Dallacasa D., Bondi M., 2002, NewAR, 46, 287

\bibitem[Stanghellini et al. 1993]{stanghellini93} Stanghellini C.,
O'Dea C.P., Baum S.A., Laurikainen E., 1993, ApJS, 88, 1

\bibitem[Str\"uder et al. 2001]{struder01} Str\"uder L., Briel U., Dannerl K., et al., 2001, A\&A 365, L18

\bibitem[Tavecchio et al. 1998]{tavecchio98} Tavecchio F., Maraschi
L., Ghisellini G., 1998, ApJ, 509, 608

\bibitem[Tavecchio et al. 2000]{tavecchio00} Tavecchio F., Maraschi
L., Sambruna R.M., Megan Urry C., 2000, ApJ, 544, L23

\bibitem[Turner et al. 1997]{turner97} Turner T.J., George I.M., Nandra K., Mushotzky R.F., ApJ 488, 164

\bibitem[Turner et al. 2001]{turner01} Turner M.J.L., Abbey A., Arnaud M., et al., 2001, A\&A 365, L27

\bibitem[Uttley et al. 1999]{uttley99} Uttley P. McHardy I.M.,
Papadakis I.E., Guainazzi M., Fruscione A., 1999, MNRAS, 307, L6

\bibitem[Zhang \& Marscher 1994]{zhang94} Zhang Y.F., Marscher A.P.,
1994, in ``Proceedings of the ROSAT Science Symposium'', Schlegel
E. \& Petre R. eds. (American Institute of Physics:New York), 406

\end{thebibliography}
\end{document}